\documentclass[aps,prd,twocolumn,superscriptaddress,showpacs,preprintnumbers,amsmath,amssymb]{revtex4}
\usepackage{graphicx}
\usepackage{dcolumn}
\usepackage{amsmath}
\usepackage{bm}

\begin{document}
\date{\today}
\title{Evidence for $B^0 \to \chi _{c1} \pi ^0$ at Belle}

\affiliation{Budker Institute of Nuclear Physics, Novosibirsk}
\affiliation{Chiba University, Chiba}
\affiliation{University of Cincinnati, Cincinnati, Ohio 45221}
\affiliation{Justus-Liebig-Universit\"at Gie\ss{}en, Gie\ss{}en}
\affiliation{The Graduate University for Advanced Studies, Hayama}
\affiliation{Hanyang University, Seoul}
\affiliation{University of Hawaii, Honolulu, Hawaii 96822}
\affiliation{High Energy Accelerator Research Organization (KEK), Tsukuba}
\affiliation{Institute of High Energy Physics, Chinese Academy of Sciences, Beijing}
\affiliation{Institute of High Energy Physics, Vienna}
\affiliation{Institute of High Energy Physics, Protvino}
\affiliation{Institute for Theoretical and Experimental Physics, Moscow}
\affiliation{J. Stefan Institute, Ljubljana}
\affiliation{Kanagawa University, Yokohama}
\affiliation{Korea University, Seoul}
\affiliation{Kyungpook National University, Taegu}
\affiliation{\'Ecole Polytechnique F\'ed\'erale de Lausanne (EPFL), Lausanne}
\affiliation{Faculty of Mathematics and Physics, University of Ljubljana, Ljubljana}
\affiliation{University of Maribor, Maribor}
\affiliation{University of Melbourne, School of Physics, Victoria 3010}
\affiliation{Nagoya University, Nagoya}
\affiliation{Nara Women's University, Nara}
\affiliation{National Central University, Chung-li}
\affiliation{National United University, Miao Li}
\affiliation{Department of Physics, National Taiwan University, Taipei}
\affiliation{H. Niewodniczanski Institute of Nuclear Physics, Krakow}
\affiliation{Nippon Dental University, Niigata}
\affiliation{Niigata University, Niigata}
\affiliation{University of Nova Gorica, Nova Gorica}
\affiliation{Osaka City University, Osaka}
\affiliation{Osaka University, Osaka}
\affiliation{Panjab University, Chandigarh}
\affiliation{Saga University, Saga}
\affiliation{University of Science and Technology of China, Hefei}
\affiliation{Seoul National University, Seoul}
\affiliation{Sungkyunkwan University, Suwon}
\affiliation{University of Sydney, Sydney, New South Wales}
\affiliation{Tata Institute of Fundamental Research, Mumbai}
\affiliation{Toho University, Funabashi}
\affiliation{Tohoku Gakuin University, Tagajo}
\affiliation{Tohoku University, Sendai}
\affiliation{Department of Physics, University of Tokyo, Tokyo}
\affiliation{Tokyo Institute of Technology, Tokyo}
\affiliation{Tokyo Metropolitan University, Tokyo}
\affiliation{Tokyo University of Agriculture and Technology, Tokyo}
\affiliation{Virginia Polytechnic Institute and State University, Blacksburg, Virginia 24061}
\affiliation{Yonsei University, Seoul}
 \author{R.~Kumar}\affiliation{Panjab University, Chandigarh} 
 \author{J.~B.~Singh}\affiliation{Panjab University, Chandigarh} 
 \author{I.~Adachi}\affiliation{High Energy Accelerator Research Organization (KEK), Tsukuba} 
 \author{H.~Aihara}\affiliation{Department of Physics, University of Tokyo, Tokyo} 
 \author{K.~Arinstein}\affiliation{Budker Institute of Nuclear Physics, Novosibirsk} 
 \author{T.~Aushev}\affiliation{\'Ecole Polytechnique F\'ed\'erale de Lausanne (EPFL), Lausanne}\affiliation{Institute for Theoretical and Experimental Physics, Moscow} 
 \author{T.~Aziz}\affiliation{Tata Institute of Fundamental Research, Mumbai} 
 \author{A.~M.~Bakich}\affiliation{University of Sydney, Sydney, New South Wales} 
 \author{V.~Balagura}\affiliation{Institute for Theoretical and Experimental Physics, Moscow} 
 \author{A.~Bay}\affiliation{\'Ecole Polytechnique F\'ed\'erale de Lausanne (EPFL), Lausanne} 
% ...
\collaboration{The Belle Collaboration}
\noaffiliation

\begin{abstract}
We present a measurement of the branching fraction for the Cabibbo- and color-suppressed $B^0 \to \chi_{c1}\pi^0$ decay based on a data sample of $657\times 10^6$~$B\overline B$ events collected at the $\Upsilon(4S)$ resonance with the Belle detector at the KEKB asymmetric-energy $e^+e^-$ collider. We observe a signal of $40\pm9$ events with a significance of $4.7\sigma$ including systematic uncertainties. The measured branching fraction is $\mathcal {B}( B^0 \to \chi_{c1} \pi^0) = (1.12\pm 0.25(\rm {stat.})\pm 0.12({\rm syst.}))\times 10^{-5}$.
\end{abstract}

\maketitle
The decay $B^0 \to \chi_{c1} \pi^0$ is a $b \to c\bar c d$ transition that proceeds at leading order through the color-suppressed tree diagram as shown in Fig. 1. If the tree diagram dominates, 
then the time-dependent $CP$-violating
asymmetries in this decay mode are predicted to be the same as
those measured in $b \to c\bar cs$ decays, such as $B^0 \to J/\psi K^0_S$~\cite{new2}. 
A deviation of the $CP$-violating asymmetries in  $B^0 \to \chi_{c1} \pi^0$ from these expectations could indicate non-negligible contributions from a penguin amplitude or amplitudes from new physics. 
For a similar $B$ decay mode, $B^0 \to J/\psi \pi^0$, 
the time-dependent $CP$-violation parameters have been measured 
by the Belle~\cite{belle_psipi0} and BaBar~\cite{babar_psipi0} 
collaborations. 
Comparison of the properties of $B^0 \to J/\psi \pi^0$ and $B^0 \to \chi_{c1} \pi^0$ decays will also provide an opportunity to probe new physics that predicts
different couplings to left-handed and right-handed particles~\cite{miya}.

The $B^0 \to \chi_{c1} \pi^0$ decay has not been observed so far. Confirming its existence is a very 
important step toward detailed studies of the $b \to c \bar c d$ 
transition. The factorization approach~\cite{one} and isospin symmetry imply 
that the branching fraction of the $B^0 \to \chi_{c1}\pi^0$ decay mode 
should be one half 
of that for $B^{\pm} \to \chi_{c1} \pi^{\pm}$~\cite{two}. 
Precise measurement of the branching fractions of these decays
can also provide information related to the final state interactions
in $B$ decays.

In this paper, we report the first 
evidence of $B^0 \to \chi_{c1} \pi^0$ using a data sample containing 
$(657 \pm 9) \times ~ 10^6~B\overline B$ events collected at the 
$\Upsilon(4S)$ resonance with the  Belle detector~\cite{belle} at 
the KEKB asymmetric-energy $e^+e^-$ collider~\cite{kekb}. The Belle detector is a large solid-angle magnetic spectrometer located at the KEKB $e^+ e^-$ storage rings, which collide 8.0 GeV electrons with 3.5 GeV positrons producing a center-of-mass (CM) energy of 10.58 GeV, the mass of the $\Upsilon (4S)$ resonance.
\begin{figure}
\begin{tabular}{c}
\includegraphics[width=0.4\textwidth]{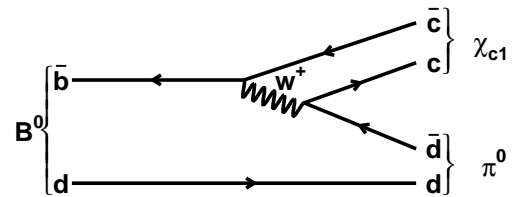}
\end{tabular}
\caption{Leading tree level diagram for $B^0 \to \chi_{c1}\pi^0$ decay.}
\end{figure}

The Belle detector consists of a silicon vertex detector (SVD), which is
surrounded by a 50-layer central drift chamber (CDC), 
an array of aerogel Cherenkov counters (ACC),
a barrel-like arrangement of time-of-flight (TOF) scintillation counters, 
 and an electromagnetic calorimeter (ECL) comprised of CsI(T$l$)
 crystals. 
These subdetectors are located inside a superconducting solenoid coil
 that provides a 1.5~T magnetic field. An iron flux-return yoke (KLM) located outside the coil is instrumented to detect $K^0_L$ mesons and to identify muons. The detector is described in detail elsewhere~\cite{belle}. The data set used in this analysis consists of two subsets: the first $152 \times 10^{6}~$ $B$ meson pairs were collected with a 2.0 cm radius beam pipe and a 3-layer SVD, and the remaining $505 \times 10^{6}~$ $B$ meson pairs with a 1.5 cm radius beam pipe, a 4-layer SVD and a small-cell inner drift chamber~\cite{yoshi,yoshi1}.

Events with $B$ meson candidates are first selected by applying the following general selection criteria for hadronic events: at least three charged tracks are required to originate from an event vertex which is consistent with the interaction point (IP); the reconstructed CM energy should satisfy $E^{CM} > 0.2 \sqrt{s}$, where $\sqrt{s}$ is the total CM energy; the component of momentum along the beam direction ($z$-direction) must be in the range $|p^{CM}_z| < 0.5 \sqrt{s}/c$; and the total ECL energy should consist of at least two energy clusters and satisfy $0.1\sqrt{s} < E^{CM}_{ECL} < 0.8\sqrt{s}$. To suppress continuum background dominated by two-jet-like $e^+e^- \to q\bar q~{\rm annihilation}~( q = u,d, s) $, we reject events where the  
ratio of the second to zeroth Fox-Wolfram moments~\cite{Fox78} $R_2$ is greater than
0.5. We find no contribution from continuum background after applying this cut on $R_2$. To remove tracks of charged particles that are poorly
measured or do not originate from the interaction region, we require their origin to be within $0.5~\rm cm$ of the IP in the radial direction, and $5~\rm cm$ in $z$-direction.
\begin{figure}
\resizebox{8.8cm}{7cm}{\includegraphics{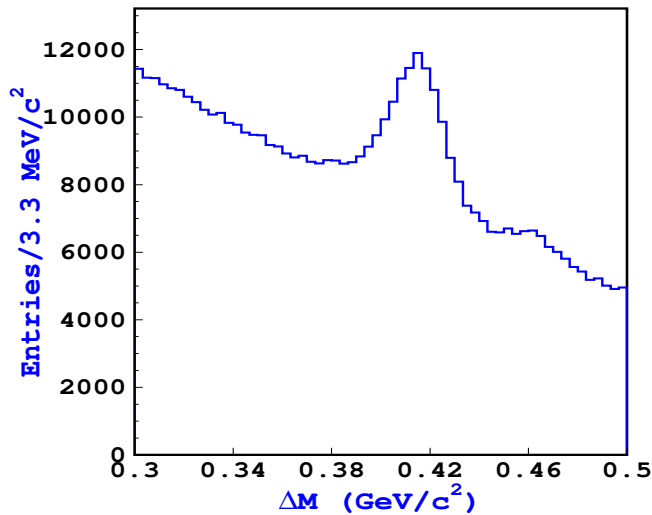}}
\caption{The $\Delta M$ ($M_{\ell ^+ \ell ^-\gamma} - M_{\ell^+\ell^-}$) distribution for inclusive $\chi _{c1}$ candidates. The enhancement just above the $\chi_{c1}$ mass region is due to the $\chi_{c2}$.}
\end{figure}

We reconstruct $J/\psi$ from pairs of $e^+e^-$ or $\mu^+\mu^-$ candidates. For muon tracks, identification is based on track
penetration depth and the hit pattern in the KLM system. 
Electron tracks
are identified by a combination of $dE/dx$ from the CDC, $E/p$
($E$ is the energy deposited in the ECL and $p$ is the momentum measured
by the SVD and the CDC), light yield in the ACC, shower shape in the ECL and position matching between an ECL cluster and extrapolated track. In order to
recover di-electron events in which one or both electrons 
radiate a photon, the four-momenta of all photons that lie within 0.05
radians of the $e^+$ or $e^-$ directions are included in the invariant
mass calculation. 
The invariant mass window used to select $J/\psi$ candidates in the $\mu ^+ \mu ^-(e ^+ e ^-)$ channel is 
$-0.06~(-0.15)~{\rm GeV/}c^2~\le M_{\ell^+\ell^-} - m_{J/\psi} \le 0.036~{\rm GeV/}c^2$, where $m_{J/\psi}$ denotes the nominal $J/\psi$ mass~\cite{Hag}; these intervals are asymmetric in order to include part of the radiative tails. 
 Vertex- and mass-constrained kinematic fits are performed for selected $J/\psi$ candidates to improve the momentum resolution. 

Photons are identified as ECL energy clusters that are not associated with a charged track and  have a minimum energy of $0.060~{\rm GeV}$. 
We reject the 
photon candidate if the ratio of the energy deposited in the array of the central $3\times 3$ ECL cells to that in the array of $5\times 5 $ cells is less than 0.82. Neutral pion candidates  are formed from the photon pairs that have an invariant mass  in the range 0.118~GeV/$c^2$ to 0.150~GeV/$c^2$. To reduce combinatorial background, an energy asymmetry threshold $|(E_{\gamma 1} -E_{\gamma 2})/(E_{\gamma 1} + E_{\gamma 2})| < 0.8$ is used, where $E_{\gamma 1}$ and $E_{\gamma 2}$ are the energies of each photon of the $\pi^0$ candidates in the CM frame. Finally, a mass constrained fit is applied to the $\pi^0$ candidates.

To reconstruct the $\chi_{c1}$ meson, we combine a $J/\psi$ candidate with momentum below 2.0~GeV/$c$ in the CM frame with a selected photon. 
To suppress  photons originating from $\pi^0 \to \gamma \gamma$, we veto photons that, when combined with another photon in the event, satisfy $0.110~{\rm GeV/}c^2 \le M_{\gamma\gamma}\le 0.150~{\rm GeV/}c^2$. 
The $\chi_{c1}$ candidates are selected by
requiring the mass difference ($\Delta M~=~M_{\ell ^+ \ell ^-\gamma} - M_{\ell^+\ell^-}$) to lie between $0.3~{\rm GeV/}c^2$ and $0.5~{\rm GeV/}c^2$. The $\Delta M$ distribution is shown in Fig. 2. A mass-constrained fit is applied to $\chi _{c1}$ candidates in order to improve the momentum resolution. 

We reconstruct $B$ mesons by combining a $\chi_{c1}$ candidate with a neutral pion. The energy difference, $\Delta E \equiv E_{B}^*- E_{\rm beam}^*$ and the mass difference ($\Delta M = M_{\ell^+\ell^-\gamma} - M_{\ell^+\ell^-}$) are used to separate signal from background, where $E_{\rm beam}^{*}$ and  $E_{B}^*$ are the run-dependent beam energy and reconstructed energy of the $B$ meson candidates in the CM frame, respectively. For the selected $B^0$ candidates, the beam-constrained mass, $M_{\rm bc} \equiv \sqrt{E_{\rm beam}^{*2} - p_B^{*2}}$, where $p_{B}^*$ is the reconstructed momentum of the $B$ meson candidates in the CM frame, is required to be $5.27~{\rm GeV}/c^2 < M_{\rm bc} < 5.29~{\rm GeV}/c^2 $. We retain $B^0$ candidates with $0.3~{\rm GeV}/c^2 < \Delta M < 0.5~{\rm GeV}/c^2$ and $-0.2~{\rm GeV} < \Delta E < 0.2~{\rm GeV}$ for the final analysis.
 The selection criteria are determined by optimizing the figure of merit, $S/\sqrt(S+B)$, where $S~(B)$ is the number of signal (background) events in the signal region ($-0.09~{\rm GeV}<\Delta E <0.05~{\rm GeV}$ and $0.380~{\rm GeV}/c^2<\Delta M <0.435~{\rm GeV}/c^2$). We assume the signal branching fraction is half of that for $B^{\pm} \to \chi_{c1} \pi^{\pm}$~\cite{two}.

The backgrounds are dominated by the $B\overline B$ events with a $J/\psi$ 
in the final state, where the $J/\psi$ is produced either directly
from $B$ decay or from the $\chi_{c1} \rightarrow J/\psi \gamma$ decay 
chain.  
We study these backgrounds using a large Monte Carlo (MC) sample~\cite{evtgen} corresponding to 3.86$\times$10$^{10}$ generic $B\overline B$ decays that includes 
all known $B \to J/\psi X$ processes and those where $B$ decays into higher charmonium states ($\chi_{c1}$, $\chi_{c2}$ or $\psi'$) that subsequently produce
$J/\psi$ in the final state.
The dominant contribution comes from 
$B^0 \to J/\psi K^0_S (\to \pi^0 \pi^0)$, 
$B^{\pm} \to J/\psi K^{\ast \pm}(892)$, 
$B^0 \to J/\psi K^{\ast 0} (892) $, 
$B^0 \to \chi_{c1} K_S^0 (\to \pi^0 \pi^0)$, and a 
few other exclusive $B \to J/\psi (\chi_{c1})+ X$ decay modes. To suppress neutral pions from $K^0_S \to \pi^0 \pi^0$ decays, 
we veto $\pi^0$s that, when combined with another $\pi^0$ in the event, 
satisfy 0.469~GeV/$c^2 < M_{\pi^0 \pi^0} < $ 0.526~GeV/$c^2$.
The $K^0_S$ veto reduces the background by 16.2\% with a signal loss of 4.2\%.
We further reduce the $B \to J/\psi K^0_S ( \to \pi^0 \pi^0)$ background 
by requiring $\cos\theta_{\rm hel} > -0.88$, 
where $\theta_{\rm hel}$ is the angle between  
the direction opposite to the $B$ momentum and the $\gamma$ direction 
in the $\chi_{c1}$ rest frame. After this requirement, we find that there is no peaking background in the $\Delta E$ signal region.
 The background from $B \to \chi_{c1}+X$ decay modes, such as $B^0 \to \chi_{c1} K_S^0 (\to \pi^0 \pi^0)$, $B^{\pm} \to \chi_{c1} K^{\ast}$, $B^{\pm} \to \chi_{c1}\rho^{\pm}$ and $B^0 \to \psi(2S)(\to \chi_{c1}\gamma)\pi^0$, forms a peak
in the $\Delta M$ signal region (called peaking background), 
while all other components are flat in $\Delta M$ (called combinatorial background). The background from $B \to \chi_{c2}+X$ decay modes is negligibly small and ignored. 

The signal yield is extracted by maximizing a two-dimensional (2D) extended likelihood function,
\begin{equation}
\mathcal{L}=\frac{e^{-\sum_{k}N_{k}}}{N!}\prod_{i=1}^{N}\left [\sum_{k}N_k\times P_k(\Delta E^i,\Delta M^i)\right ],
\end{equation}
where $N$ is the total number of the candidate events, $i$ is the event number, $N_k$ and $P_k$ are the yield and probability density function (PDF) of the component $k$, which corresponds to signal, peaking background and combinatorial background. The scatter plot of $\Delta M$ versus $\Delta E$ for $B^0 \to \chi_{c1} \pi^0$ candidates is shown in Fig. 3. The number of events in the ($\Delta E, \Delta M$) fit region is 357.

The signal PDF is modeled using a  Crystal Ball (CB) lineshape function~\cite{cry} for $\Delta E$ and a CB lineshape function for $\Delta M$ whose shape parameters are determined from a signal MC sample. 
To take into account a small difference between data and MC, the shapes of the $\Delta E$ and $\Delta M$ distributions are corrected according to calibration constants obtained from the $B^+ \to J/\psi K^{\ast +} (K^{\ast +} \to K^+ \pi^0)$ and $B^- \to \chi_{c1}K^-$ samples, respectively. In the reconstruction of $B^+ \to J/\psi K^{\ast +}$ events, we require the momentum of the $\pi^0$ to be greater than 0.75~GeV/$c$. This requirement results in a $\Delta E$ distribution similar to that for the signal events. The calibration constants for the mean and width of $\Delta E$ ($\Delta M$) are found to be $-6.23\pm0.97$~MeV~($-1.16\pm0.46$~MeV) and $1.37\pm0.07$~($1.12\pm0.05$), respectively.
\begin{figure}
\resizebox{8cm}{7cm}{\includegraphics{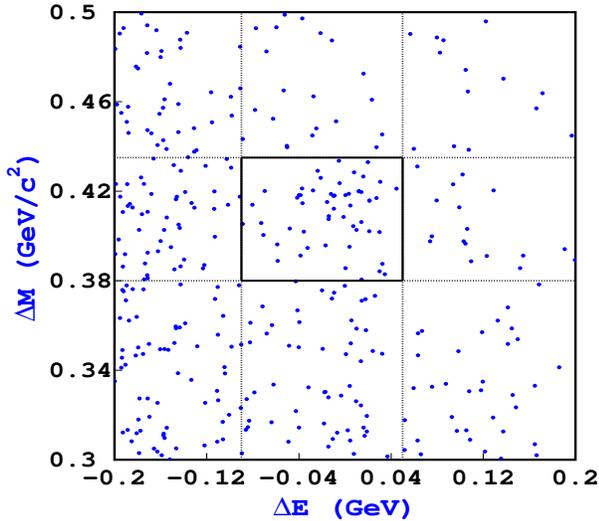}}
\caption{ Scatter plot of $\Delta M$ versus $\Delta E$ for $B^0 \to \chi_{c1} \pi^0$ candidates. The dashed lines and solid box indicate the signal region.}
\end{figure}

The peaking background shape is modeled using a CB lineshape function in $\Delta M$ and an exponential function in $\Delta E$. The shape parameters of the CB lineshape function are fixed from MC and those for the exponential function are floated. The shapes of combinatorial backgrounds are modeled by a first-order polynomial function for $\Delta E$ and a second-order polynomial function for $\Delta M$. The shape parameters and the number of combinatorial background events are allowed to float in the fit. 

The $\Delta E$ and $\Delta M$ distributions along with the  projections of the fit are shown in Fig. 4. The fit yields a signal of $40\pm9$ $B^0 \to \chi_{c1}\pi^0$ candidates. The number of peaking background events is $14\pm7$, which is in good agreement with MC expectations of $14$.
 
The significance of the $B^0\to \chi _{c1} \pi^0$ signal is 4.7$\sigma$, defined as 
$\sqrt{-2\ln(\mathcal L_0/\mathcal L_{\rm max})}$ and 
$\mathcal L_{\rm max}$ ($\mathcal L_0$) denotes the maximum likelihood value (the value obtained from the fit when signal yield fixed to zero). 
We include the effect of systematic uncertainties by subtracting the quadratic sum of the variations of the significance in smaller direction when each fixed parameter in the fit is changed by $\pm1\sigma$. 
\begin{figure}
\resizebox{8.8cm}{7cm}{\includegraphics{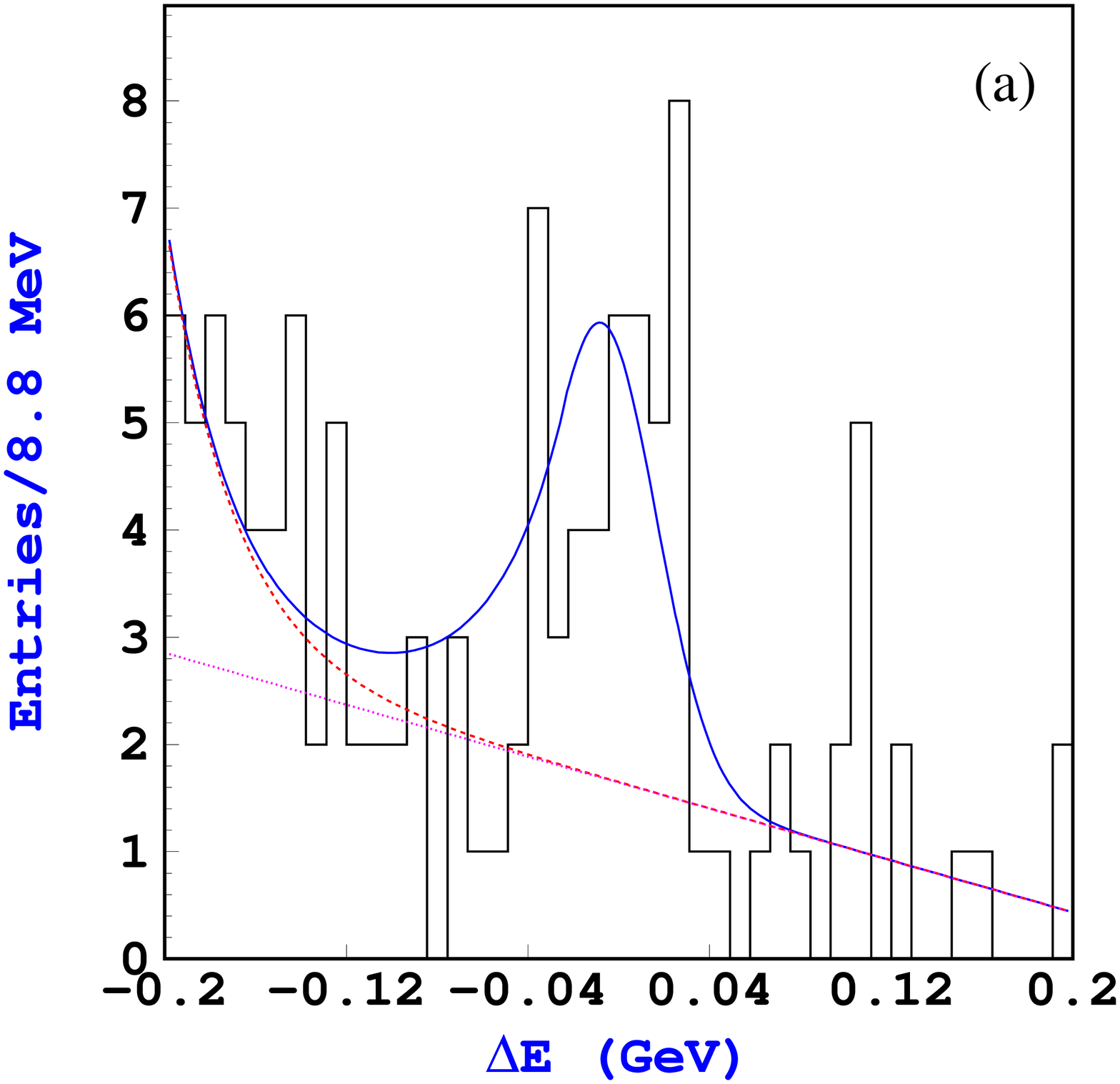}}
\resizebox{8.8cm}{7cm}{\includegraphics{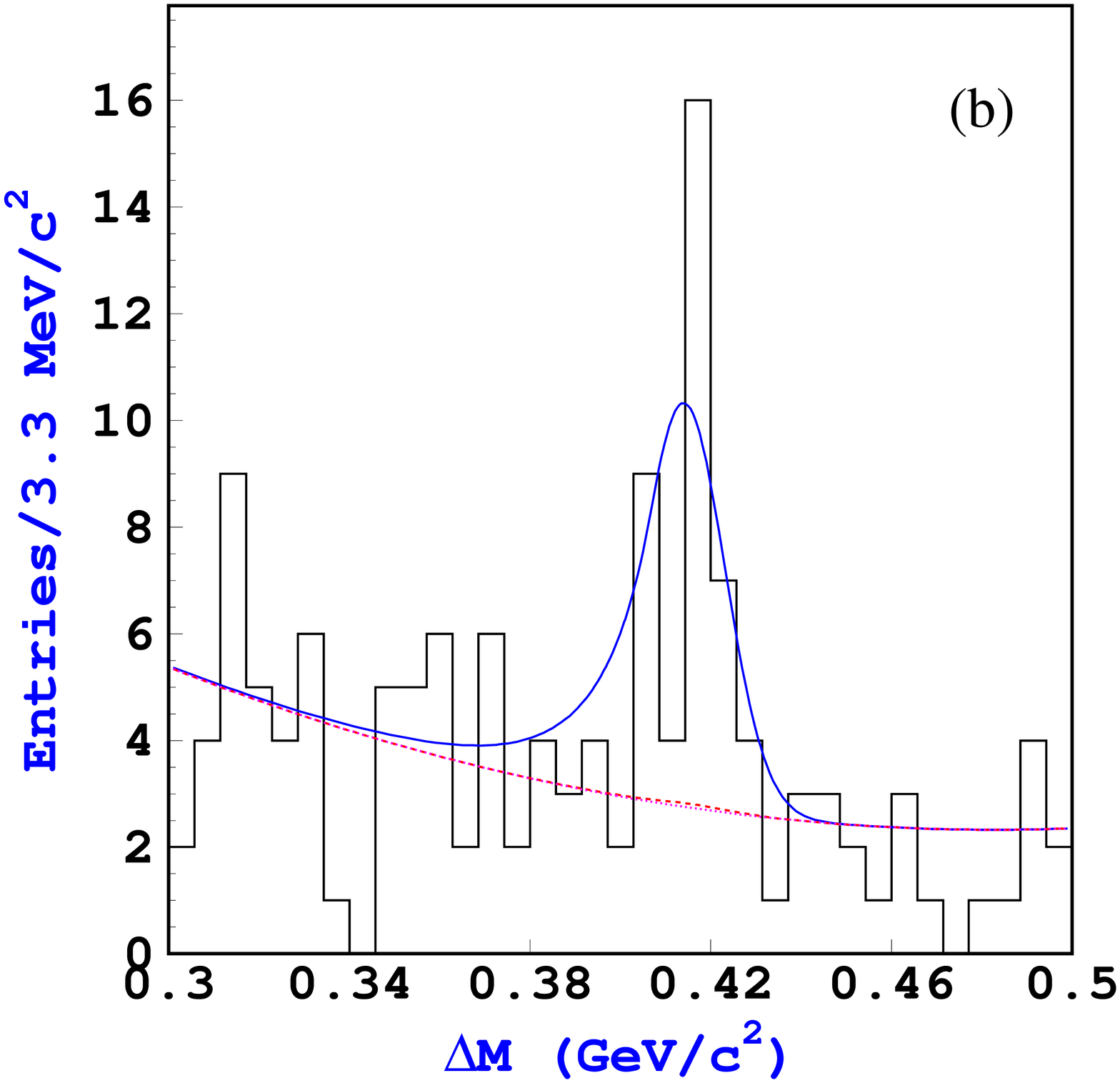}}
\caption{(a) The  projections in  $\Delta E$ for events satisfying $0.380~{\rm GeV}/c^2 < \Delta M < 0.435~{\rm GeV}/c^2$ and (b) the projections in $\Delta M$ for events satisfying $-0.09~{\rm GeV} < \Delta E < 0.05~{\rm GeV}$. The solid curve represents the overall fit, dashed curve represents the sum of peaking and combinatorial background, and dotted curve only combinatorial background component.}
\end{figure}
The branching fraction ($\mathcal{B}$) for the $B^0\to {\chi}_{c1}\pi^0$ decay mode is calculated as follows: 
\begin{equation}
\mathcal{B} = \frac{N_{\rm sig}}{\epsilon \times N_{B\overline B}\times {\mathcal {B}_{\rm sec}}},
\end{equation} 
where $N_{\rm sig}$ is the observed signal yield, $\epsilon$ ($N_{B\overline B}$) is the reconstruction efficiency (number of $B$ mesons in the
 data sample), and $\mathcal{B}_{\rm sec}$ is the product of $\mathcal{B}(\chi_{c1}\to J/\psi \gamma)$, $\mathcal{B}(J/\psi\to \ell \ell)$ and $\mathcal{B}(\pi^0 \to \gamma \gamma)$. The reconstruction efficiency is determined from a signal MC sample. After a small correction for the muon identification requirement the efficiency is found to be 13.0\%. We use the daughter branching fractions published in Ref.~\cite{Hag}. 
Equal production of neutral and charged $B$ meson pairs in
 $\Upsilon ({\rm 4S})$ decay is assumed. 
The resulting branching fraction is  
\begin{equation}
  \mathcal {B}( B^0 \to \chi_{c1} \pi^0) 
     = (1.12\pm 0.25\pm 0.12)\times 10^{-5},
\end{equation}
where the first error is statistical and the second is systematic.

The systematic uncertainties are summarized in Table~\ref{tab:table2}.
The systematic uncertainty on the signal yield is calculated by varying each shape parameter fixed in the fit by $\pm$1$\sigma$, and then taking the quadratic sum of the deviations from the nominal value.  
We have checked for possible bias in the fitting using a MC sample; no significant bias was found. The systematic uncertainty assigned to the signal yield is 4.5$\%$.
The uncertainty on the tracking efficiency is estimated to be 1.0$\%$ per track, while that due to lepton identification is $3.9\%$. 
We also assign an uncertainty of 4.1\% for $\pi^0 \to$ $\gamma \gamma$ reconstruction, and  an uncertainty of $2.0\%$ for the $\gamma$ detection efficiency; these are correlated and added linearly (6.1\%). 
The systematic uncertainty due to the ${\chi}_{c1}\to \gamma J/\psi$ and $J/\psi\to \ell^+ \ell^-$ branching fractions is 5.5$\%$.
The total systematic error is the sum of all the above uncertainties in quadrature.

\begin{table}
\caption{\label{tab:table2} Summary of systematic errors on
   branching fraction.   
}
\begin{ruledtabular}
\begin{tabular}{lc}                      
 Source & Uncertainty (\%)
\\\hline
Yield uncertainty & 4.5  \\         
Tracking  & 2.0         \\         
Lepton identification & 3.9             \\
$\gamma$ and $\pi^0$ detection & 6.1 \\ 
$\pi^0$ veto & 1.6\\          
MC statistics & 2.3         \\          
$\rm {N}_{B\bar B}$ & 1.4    \\  
Secondary branching fractions  & 5.5     
\\\hline
Total & 10.8            \\          
\end{tabular} 
\end{ruledtabular}
\end{table}
In summary, we report the first evidence of $B^0 \to \chi_{c1}\pi^0$ with $(657\pm9)\times10^6$~$B\overline B$ events. The observed signal yield is $40\pm9$ with a significance of 4.7$\sigma$ including systematic uncertainty. The measured branching fraction is $\mathcal {B}( B^0 \to \chi_{c1} \pi^0) = (1.12\pm 0.25\pm 0.12)\times 10^{-5}$, which is consistent with the factorization model.

We thank the KEKB group for excellent operation of the
accelerator, the KEK cryogenics group for efficient solenoid
operations, and the KEK computer group and
the NII for valuable computing and SINET3 network
support.  We acknowledge support from MEXT and JSPS (Japan);
ARC and DEST (Australia); NSFC (China); 
DST (India); MOEHRD, KOSEF and KRF (Korea); 
KBN (Poland); MES and RFAAE (Russia); ARRS (Slovenia); SNSF (Switzerland); 
NSC and MOE (Taiwan); and DOE (USA).


\end{document}